# A damage model for granite subjected to quasi-static contact loading

H. Shariati [a], M. Saadati [a,b,*], K. Weddfelt [b], P.-L. Larsson [a], F. Hild [c]

[a] *Department of Solid Mechanics, KTH Royal Institute of Technology, Stockholm, Sweden*

[b] *Epiroc Rock Drills AB, Örebro, Sweden*

[c] *Université Paris-Saclay, ENS Paris-Saclay, CNRS, LMT–Laboratoire de Mécanique et Technologie, Gif-sur-Yvette, France*

**Abstract**

An anisotropic damage model is employed in order to simulate the fracture pattern of Bohus granite under quasi-static spherical indentation loading. The chosen damage description is added to the previously employed Drucker-Prager plasticity model with variable dilation angle. The resulting constitutive model is implemented to simulate the behavior of Bohus granite under indentation up to the load capacity of the material. The initial fragmentation, corresponding to the first small load-drop in the force-penetration curve, is likely due to the high radial tensile stress state at or close to the contact boundary. Both predicted fracture pattern and force-penetration results from the numerical simulation are compared to experimental data and a good agreement is found. The variability in tensile strength of the material is included in the chosen damage model by taking advantage of Weibull statistics. In this work, it is suggested that indentation test results by themselves may be used to calibrate the statistical distribution of tensile strength at small size scales such as in indentation applications.

## 1 Introduction

The fracture pattern of quasi-brittle materials under quasi-static (Q-S) or dynamic loading conditions is investigated extensively in the literature. In one of the first studies [1], a number of circular flat-bottomed indenters with various sizes were used to perform Q-S indentation tests on rocks. The force-penetration ($P$-$h$) response was recorded to describe the different stages of the fragmentation process that consisted of the formation of a crater, a crushed zone and a region with various cracks. Taking advantage of numerical modeling, a rock and tool interaction code was developed to simulate the fragmentation process during loading of such materials by single and double indenters [2]. A double elliptic strength criterion representing a transition from brittle to ductile failure with increasing confining pressure was considered. In another study [3], isotropic damage was used to numerically simulate the fragmentation of rock due to indentation. 2D plane strain conditions were assumed and inelasticity was ignored in both mentioned studies. Moreover, an isotropic damage model for tensile loading conditions was combined with a viscoplastic model and a cap for compressive loading conditions [4]. The employed model was utilized to simulate the rock behavior under low-velocity impact.

---

* Corresponding author: Department of Solid Mechanics, KTH Royal Institute of Technology, Stockholm, Sweden.
*Email address*: msaadati@kth.se (M. Saadati).

The fracture pattern of indented Bohus granite was investigated in the literature. In one study [5], its fragmentation induced by percussive drilling was analyzed. The (DFH) anisotropic damage model proposed by Denoual et al. [6,7] was combined with the (KST) plasticity model introduced by Krieg et al. [8,9] in order to predict the fracture pattern of the material. In another study [10], the fracture mechanism of the same rock (Bohus granite) was investigated when indented by a circular flat punch using the theory of elasticity. The focus was mainly put on the final load capacity of the selected rocks, which is due to material cracking features. The model parameters in both mentioned studies were calibrated against experiments including direct tension and compression, three-point bending and quasi-oedometric tests [11]. The tensile strength distribution was described by Weibull statistics [12,13], and the corresponding parameters were obtained from three-point bend tests. Moreover, it is well-known that the tensile strength of rocks is a size-dependent variable [14]. Therefore, the Weibull size effect was adopted in both studies [5,10] and the rock tensile strength determined from three-point bend tests was scaled using the effective volume, which was obtained analytically.

The inelastic behavior of Bohus granite under Q-S spherical indentation loading was also investigated [15]. In that study, any kind of inelastic deformation except tensile failure (i.e., mode I fracture) was described within the theory of plasticity. Quasi-oedometric compression tests were utilized to calibrate the constitutive model. It was suggested that the linear Drucker-Prager law could be used to simulate the behavior of the material. The corresponding dilation angle was determined as a function of hydrostatic pressure. Then, the calibrated constitutive model was validated by simulations of spherical indentation. It was attempted to predict the force-penetration response of the material in indentation tests up to the main load drop. Moreover, a high speed camera was utilized during some of the indentation tests. The acquired images showed that each load-drop in the *P-h* response corresponded to a material removal occurrence on the specimen surface. It was elaborated that these load-drops were possibly due to ring cracking.

It should be emphasized that, at contact, the size of the plastic zone is small compared to, for example, the corresponding metallic situation, as with the material properties relevant to granite the contact behavior falls within the so-called elastoplastic regime [16]. In this regime, elastic strains are of the same order of magnitude as plastic strains. For this situation, Alehossein et al. [17] detailed the cavity expansion model suggested by Johnson [16] to include constitutive features suited for rock materials.

In the following, it is attempted to add a damage description to the constitutive model discussed above [15]. In the referred paper, the *P-h* response of Bohus granite under Q-S indentation was numerically predicted only using the plasticity theory. Thus, tensile failure was not included in that study. In this work, the DFH anisotropic damage model is used in order to account for tensile failure in the constitutive model. Adding the damage model is necessary to determine the first stages of cracking, and to obtain local field variables in an accurate manner. During the indentation test, the rock material experiences high tensile stresses on the surface, outside the contact region. As was already shown [15], there were some small load drops in the *P-h* response of the material during indentation tests. Each load drop corresponded to local material failure on the surface. It was stated that the tensile stress level on the surface exceeded the tensile strength of the material, and consequently, the material failed locally. The stress level corresponding to the first load drop was considered as the first instance for which the tensile strength of the tested rock specimens was reached.

Furthermore, the load and tensile stress levels at which the first load drop occurred changed between experiments. As a result, the tensile strength of the material should be treated as a random variable. In doing so, the Weibull theory is utilized and the Weibull modulus is determined. The Weibull size effect is considered and the effective volume is calculated based on the stress state during the first load drop. Only regions with positive principal stresses are accounted for in effective volume calculations. The constitutive parameters including those describing damage, namely, the Weibull modulus and the effective volume, as well as the previously determined plastic parameters [15] are selected in order to numerically simulate indentation tests. As a result, a localized damage zone on the rock surface is numerically predicted and compared with experimental observations provided by a high speed camera [15]. Moreover, the damage description effect is studied on the predicted *P-h* response of the material. Statistical effects related to the damage model are also investigated.

## 2  Inelastic constitutive model (theory of plasticity)

In a previous study [15], quasi-oedometric (QO) test results were used to determine a constitutive model for Bohus granite. The equivalent stress was plotted against the pressure for QO test results, and a linear Drucker-Prager (DP) yield function was fitted. The dilation angle was also determined as a function of pressure (Figure 1). At low pressures, the interpretation of the experimental results is delicate due to low plastic strain levels and localization issues, but is adhered to presently as they were consistent with those at higher pressure levels. The linear DP criterion is formulated as

$$F = q - p \tan\beta - d = 0 \tag{1}$$

where $F$ is the yield function, $d$ the cohesion of the material, $\beta$ the friction angle, and $q$ von Mises equivalent stress. The flow potential $G$ is written as

$$G = q - p \tan\Psi \tag{2}$$

where $\Psi$ is the dilation angle, which is determined as

$$\Psi = \tan^{-1}\left(\frac{3}{3\frac{\dot{\varepsilon}_a^i}{\dot{\varepsilon}_v^i} - 1}\right) \tag{3}$$

where $\dot{\varepsilon}_a^i$ is the inelastic axial strain rate, and $\dot{\varepsilon}_v^i$ the inelastic volumetric strain rate. The flow rule is non-associated, which means that the inelastic flow occurs at angle $\Psi$ with respect to the *q*-axis of the *q-p* plane (i.e. $\Psi \neq \beta$).

It is worth mentioning that the dilation angle is a critical material parameter that should be carefully treated mainly due to the fact that it has a significant influence on the inelastic behavior [18]. It is either neglected, which results in incompressible flow rule [19], or assumed equal to the friction angle in an associated flow rule [20]. Neither of these approaches are necessarily representative of the inelastic behavior of rocks [18,21]. Moreover, in practice, measuring such parameter especially at high levels of hydrostatic pressure is difficult.

Therefore, the approximation of a constant dilation angle is usually made [4]. However, the dilation angle is believed to change as a function of inelastic strain level [22–25]. The dilation angle obtained from QO compression tests [11] was added to the plasticity model in order to account for inelastic volumetric changes due to shear. It is worth noting that the dilation angle shown in Figure 1 is very high at least initially (i.e., at low pressures). However, the experimental results are well in line with other (experimental) data [25] for granite and will accordingly be relied upon hereafter as they are validated with the present numerical effort. In the present analysis, the dilation angle should be smaller than the friction angle due to thermodynamic restrictions.

The resulting constitutive model was used to numerically simulate Q-S spherical indentation, which is schematically shown in Figure 2, and the force-penetration *P-h* response of the material was predicted. The numerical results were compared with indentation test results and a good agreement was found. Further, a high speed camera was utilized in order to investigate the reason behind the load-drops in the *P-h* response of the material during indentation test. The attention was put on the initial loading regime up to the first large load drop in the *P-h* response, thereby defining the load capacity of Bohus granite. Images from the rock specimen surface were acquired before and after each load drop. It was shown that each of them corresponded to material removal on the specimen surface. Last, the tested specimens were *a posteriori* imaged by Computed microTomography (μCT) and no visible sub-surface crack was detected. It was speculated that the small load-drops were due to ring (Hertzian) cracks immediately outside the contact area, and those cracks had negligible influence on the *P–h* response.

In this section it is worth mentioning another plasticity model used previously [11]. The authors employed the pressure-dependent KST model. The KST model is close to the DP model in spirit, but with zero dilation angle. Later in this study, both KST and DP models together with a damage description are considered as the constitutive laws to predict the *P-h* response of the material during indentation tests. The predicted results are compared with the experimental data and followed by a discussion.

## 3  Damage constitutive model

The fracture pattern and *P-h* response of Bohus granite under quasi-static spherical indentation loading condition is of interest in this study. The stress state beneath the indenter is complex. It consists of both compressive and tensile stress states. The behavior of granite due to compressive stresses is described using a linear DP criterion with variable dilation angle. A damage model is introduced to the previous model [15] in order to account for tensile failure. For this purpose, the DFH anisotropic damage model is employed. One key assumption is that initial defects with different sizes and orientations are randomly distributed within the material. Therefore, the failure stress (i.e., the tensile stress level at which the material fails) is a random variable. Consequently, the material response due to tensile stresses should be considered with probabilistic features. The Weibull model and weakest link theory are utilized for this purpose [7,12,13]. Using a Poisson point-process framework, the weakest link assumption and Weibull model, the failure probability is

$$P_F = 1 - \exp[-Z_{eff} \lambda_t(\sigma_F)] \tag{4}$$

where $Z_{eff}$ is the effective volume [26], and $\lambda_t$ the initiation density

$$\lambda_t(\sigma_F) = \lambda_0 \left(\frac{\sigma_F}{S_0}\right)^m \tag{5}$$

where $m$ is the Weibull modulus, $S_0^m/\lambda_0$ the Weibull scale parameter, and $\sigma_F$ the maximum tensile stress in the material. The effective volume is defined as

$$Z_{eff} = Z H_m \tag{6}$$

where $Z$ is the total volume, and $H_m$ the stress heterogeneity factor

$$H_m = \frac{1}{Z} \int_\Omega \left(\frac{\langle \sigma_1 \rangle}{\sigma_F}\right)^m dZ \text{ when } \sigma_F > 0 \tag{7}$$

where $\sigma_1$ is the maximum principal stress, $\sigma_F$ its maximum level in the whole considered volume, and $\langle \cdot \rangle$ denote the Macauley brackets. The average failure stress $\sigma_w$ reads

$$\sigma_w = S_0 (\lambda_0 Z H_m)^{-\frac{1}{m}} \Gamma\left(1 + \frac{1}{m}\right) \tag{8}$$

where $\Gamma$ is the Euler function of the second kind

$$\Gamma(x) = \int_0^\infty \exp(-u) u^{x-1} du \tag{9}$$

and the corresponding standard deviation becomes

$$\sigma_{sd} = S_0 (\lambda_0 Z H_m)^{-\frac{1}{m}} \sqrt{\Gamma\left(1 + \frac{2}{m}\right) - \Gamma^2\left(1 + \frac{1}{m}\right)} \tag{10}$$

For the anisotropic damage description, the initiation probability $P_{Fi}$ ($i = 1, 2, 3$) defines the first damage event (i.e. local failure) for each principal direction.

The DFH model is capable of predicting the tensile behavior of brittle materials for both low and high loading rates [27]. Under low loading rate conditions, fragmentation is due to the initiation and growth of a single crack that is created from the propagation of the weakest defect in the material. But if the loading rate is higher, the stress reaches high levels over time and multiple defects are activated. The latter ones cause initiation and propagation of several cracks and fragmentation. Crack propagation leads to stress relaxation in its vicinity and stops the activation of new defects in obscured zones centered about each propagating crack. However, in the non-obscured zone, the stress increases and new cracks may initiate due to the activation of critical defects. At some point, the whole domain is obscured by the propagating cracks and, as a result, the fragmentation process stops.

As for fragmentations, the interaction between the region around a crack, where stresses are relaxed, and the boundary of the domain is small and assuming a uniform stress around each integration point, the obscuration probability is defined as

$$P_o(T) = 1 - P_{no}(T) = 1 - \exp\left(-\int_0^T \frac{d\lambda_t}{dt}[\sigma(t)]Z_o(T-t)dt\right) \quad (11)$$

here $Z_o$ is the obscured zone, $\sigma$ the local principal stress, $T$ the current time and $t$ the crack initiation time. $P_{no}$ in Eq. (11) is the probability of non-obscuration, which gives the interaction law between the already initiated cracks and the critical defects within the material. It is worth noting that $P_o$ is defined for each principal direction, $P_{oi}$ ($i = 1, 2, 3$), provided the corresponding principal stress is positive, and corresponds to the three damage variables of the anisotropic damage model [27]. In order to employ the change of obscuration probability in the DFH implementation, each damage variable $\omega_i \equiv P_{oi}$ is expressed in a differential form as

$$\frac{d^2}{dt^2}\left(\frac{1}{1-\omega_i}\frac{d\omega_i}{dt}\right) = 3!\, S(kC_0)^3 \lambda_t[\sigma_i(t)] \quad \text{when } \frac{d\sigma_i}{dt} > 0 \text{ and } \sigma_i > 0 \quad (12)$$

where $\sigma_i$ is the local principal stress, $S$ a shape parameter (equal to $4\pi/3$ when the obscuration volume is similar to a sphere), $k$ a constant parameter (equal to 0.38), and $C_0$ the 1D wave speed (so that $kC_0$ corresponds to the crack propagation velocity). For more detailed descriptions of the DFH model, the reader may refer to Refs. [6,7]. The same DFH model was used in a previous study [5]. However, in this work it is coupled with a different plasticity model to capture the compressive response more accurately.

In a previous study [11], experiments including direct tension and compression, three-point bending and QO tests were conducted to determine the mechanical properties of Bohus granite. More specifically, the quasi-static tensile strength of the material was determined from three-point bend (3PB) tests. Specimens of size $40 \times 40 \times 150$ mm$^3$ were loaded quasi-statically up to failure. The nominal stress-strain behavior was mainly linear followed by brittle failure. The tensile strength of the material was determined using the Euler-Bernoulli beam theory. The tensile strength distribution was described by Weibull statistics, which is presently referred to as statistical effects in the damage model. The Weibull modulus $m$, identified from the Weibull plot, was equal to 23, the effective volume was 189 mm$^3$, and the mean tensile strength 18.7 MPa. However, for the indentation test, the effective volume was much lower than that of 3PB tests [10]. Therefore, the hypothesis of using the (extrapolated) scaled rock tensile strength determined from 3PB tests as the reference flexural strength in the indentation case may lead to unrealistic results as discussed later on.

## 4 Material parameters

The constitutive law used in numerical simulations was calibrated based on experimental data. It is composed of three different parts being elasticity, plasticity and damage. Each part has several parameters to calibrate. The elastic modulus and Poisson's ratio were determined from experimental data of uniaxial tensile and compressive tests [11]. The Drucker-Prager (DP) model was considered to describe inelasticity, and the required parameters were calibrated with QO test results [15]. Last, the DFH model was chosen as the damage description of the material. In practice, the constitutive model was implemented numerically via a VUMAT routine in the explicit finite element code Abaqus [28] combining the DP model with the existing DFH model. The resulting model is designated as DP-DFH throughout this paper.

Rock fractures are due to the initiation, growth and coalescence of cracks together with sliding between the grains and the crack surfaces. Cracks as well as some other microstructural features such as grains with different properties, voids and pores with different sizes and shapes are considered as the reason behind the heterogeneity in the strength of rocks [29]. Therefore, the tensile strength of the material should be treated as a random variable and may be described by Weibull statistics. Moreover, the Weibull size effect is also to be considered since granite shows a size-dependent behavior in the terms of its tensile strength. For instance, in the study of Weddfelt et al. [10], the fracture mechanism of the same rock (Bohus granite) indented by a circular flat punch was investigated. The effective volume was calculated and used to scale the rock tensile strength determined from the 3PB tests. The tensile strength of the material for indentation loading, as a result of the Weibull size effect, is proportional to the tensile strength of the material determined from 3PB tests corrected by the factor $\left(V_{eff,3PB}/V_{eff,indentation}\right)^{1/m}$.

In the present study, it is first attempted to use the material properties obtained from 3PB tests, namely $\sigma_w = 18.7$ MPa, $V_{eff} = 189$ mm$^3$ and $m = 23$ for Bohus granite as reference values in the Weibull size effect. However, the effective volume in indentation tests is two orders of magnitude smaller than that in the mentioned 3PB tests [10]. In such a small volume, which is of the order of the grain size in this material, the statistical behavior may be different from 3PB tests. This is due to the fact that the material may be considered as more heterogeneous at the effective scale of indentation tests in comparison to larger scales. Therefore, either a very small size 3PB test should be performed, which is not practical, or another (closer) condition to the indentation test should be considered.

Another route is followed herein. It is suggested, as also proposed by Olsson et al. [30], that the indentation test results *by themselves* may be used to calibrate the statistical parameters of the model at such small effective scales. The first load-drop in the *P-h* response of the material in indentation tests is considered to determine the Weibull parameters. The first load drop is likely due to high radial tensile stresses at or close to the contact boundary. In practice, the stress state during the initial steps of the indentation test is determined based on finite element simulations.

The experimental data from Q-S spherical indentation tests are used to determine the damage parameters, namely, the Weibull modulus and the effective volume. Specifically, the recorded *P-h* response during a number of indentation tests [15] are considered. The "experimentally determined force-penetration response of Bohus granite for spherical indentation tests" (Figure 4 in Ref. [15]) only shows 4 experimental responses for the sake of clarity. More experimental data were available, which are used to determine the Weibull modulus *m*. It is seen in the afore-mentioned figure that the initial part of the *P-h* response is almost deterministic. Therefore, the *P-h* response of the material was captured using the theory of plasticity (i.e., DP model). Furthermore, it was speculated, as previously mentioned, that if the maximum tensile stress level right outside the contact zone reached the tensile strength of the material, it failed and as a result of this failure, the load level dropped. Accordingly, the stress state corresponding to the first load drop can possibly be seen as the indentation tensile strength of the tested specimens being reached. However, the load drops, and specifically the first load drops that are of interest in this study, occur at somewhat different indentation depths. Therefore, the tensile strength of the material should be considered as a random variable.

In practice, the way to determine the Weibull modulus $m$ is to plot $\ln(-\ln(1-P_f))$ vs. $\ln(\sigma_f)$ where $P_f$ is the failure probability, and $\sigma_f$ the tensile failure stress [31]. To obtain the so-called Weibull plot, $N = 12$ indentation experiments were considered, and the corresponding indentation depth at the first load-drop occurrence was determined. Then, the maximum tensile stress level at that indentation depth was obtained from finite element simulation results. It is worth noting that the simulation results are not affected by the previously discussed statistical effects in the damage model as before the first small load-drop in the force-displacement curve no damage occurred yet (i.e., the indentation load was still rather low). The resulting failure stresses were sorted in ascending order. Moreover, an estimate of the failure probability $P_f$ of the $n$-th test was calculated as

$$P_f = \frac{n - 0.5}{N} \tag{13}$$

Last, a linear regression was performed, see Figure 3. The slope of the fitted line yields the Weibull modulus $m$. According to the results shown in Figure 3(a), the $m$-value given when considering all data points is 6. However, three of the test results had a very low tensile strength and when including these points, a true Weibull behavior is not found. This may be due to the presence of two populations of defects in the material or inaccuracies, for example, when determining the first load drop. For this reason, these three points were disregarded and the $m$-value becomes equal to 24. This result is in line with the $m$-value previously determined [11]. Furthermore, the effective volume is determined using Equations (6) and (7), and is equal to 1 mm³ when only elements with positive maximum principal stresses are considered. The material parameters for Bohus granite, as modeled presently, are summarized in Table 1.

## 5 Finite Element Model

The details of the finite element model can be found in Ref. [15]. Here a brief summary is presented. The explicit version of Abaqus [28] is used for numerical simulations. Taking advantage of symmetries, but remembering that the tensile strength is not axisymmetrically distributed, only one quarter of rock specimen is modeled. Eight-node linear reduced integration brick elements with 0.2 mm in length are used to mesh the rock geometry. The indenter is modeled as a 3D analytical rigid body since its Young's modulus is ten times larger than that of Bohus granite. The load is applied in a displacement control mode by adding a vertical displacement to the indenter reference point. The mesh is shown in Figure 4. It should be mentioned that the contact behavior is assumed to be frictionless.

The material model was implemented in the commercial code Abaqus explicit as a VUMAT subroutine. The DP constitutive model was included to the existing DFH code. Then, the resulting DP-DFH computational code was used to simulate indentation tests. All the material parameters accounting for elasticity, plasticity and damage are defined as inputs to the DP-DFH subroutine. It is worth emphasizing that a continuum mechanics approach was relied upon in the finite element simulations reported herein.

The constitutive (DP-DFH) model implementation through the Abaqus VUMAT subroutine is briefly explained here. First, the total solution time is divided into very small-time steps by Abaqus. At the beginning of each time step, the strain and stress states are known from the previous step in addition to all internal variables. The strain increments are then introduced by Abaqus and consequently the stress state should be updated. The trial (predicted) stress state is determined assuming pure elastic deformation. This stress state is then "corrected" with respect to the yield function whenever necessary and plastic strains are updated. The principal stresses are found based on the "corrected" stress state. As soon as any principal stress ($\sigma_i$) reaches the material tensile strength, the corresponding damage variable ($\omega_i$) will grow. As the solution proceeds, the damage variables evolve based on the growth law (12). The damage variable can acquire any value ranging from zero, corresponding to intact material, to one, which means that the material cannot hold any tensile load in the corresponding direction ($0 < \omega_i < 1$). Finally, the stress state is updated by multiplying each (effective) principal stress $\sigma_i$ by $(1 - \omega_i)$. A schematic flowchart of the numerical implementation is shown in Figure 5.

The statistical effects included in the damage model are also investigated. Ten subsequent simulations were performed, and quantities such as crack pattern and plastic zone size were studied.

## 6 Numerical Results and Discussion

The force-penetration response from Experiment 1 [15] as well as those from the numerical simulations are plotted in Figure 6. The DP, DP-DFH, KST and elastic models are compared with experimental results. For the KST model parameters, the reader may refer to Ref. [11]. The material parameters corresponding to the DP-DFH model are gathered in Table 1. The results for DP and DP-DFH models are in very close agreement with the experimental results. When comparing them with the corresponding KST results, it is concluded that the DP model introduces effects from compressible plasticity, which are in agreement with the mechanics of the problem. The effects from the damage model are discussed next.

It is first attempted to use the DFH material properties namely $\sigma_w = 18.7$ MPa, $V_{eff} = 189$ mm$^3$, and $m = 23$ for Bohus granite [11]. The resulting constitutive model numerically predicts large crack planes. These crack planes extend deep into the material. However, these cracks were not visible in μCT images [15].

If the DP-DFH parameters of Table 1 are used in the calculations, the damage region shown in Figure 7 is observed. The radius of the formed crater on indented specimens (Figure 8(a)) is in good agreement with the numerical simulation results (Figure 8(b)). The simulation result for the fracture pattern does not appear to be the same as the crater formed after the major load drop in the indentation test. The fragmentation process in this very brittle material starts with ring cracks at early loading without total failure of the material (these cracks are observable from the simulation results if they are plotted at early loading) and as the loading proceeds, more tensile stresses are built within the material especially close to the free surface. It is hypothesized that the major load drop is triggered by the radial cracks shown in Figure 8(b), and that the final fracture of the material leading to multiple fragmentation is governed by the elastic energy stored within the material. The current model is not able to provide the details of that fragmentation response during and after the major load drop.

In a homogenous elastoplastic material description and in the absence of any damage model, the numerical simulation predicts a region with both positive first and mid principal stresses around the contact area (Figure 9(a-d)). Opposite to the Hertzian case, the direction of the first principal stress becomes circumferential around the contact area, except immediately outside the contact boundary, which is mainly due to plastic deformations. This stress leads to the formation of radial cracks in the numerical simulation when including the damage model (Figure 7). The third principal stress, as expected, is compressive and normal to the contact surface along the axis of symmetry (see Figure 9(e-f)).

Further investigations of the statistical effects in the damage model were performed. Ten simulations with different realizations of the tensile strength were conducted. The outcome of these analyses concerning the damage pattern is shown in Figure 10(a) (extreme results are displayed). The size of radial cracks is virtually independent of the local statistical details. Corresponding results for the plastic zone size, Figure 10(b), and field variables show that these quantities were essentially not affected by the statistics included in the damage model.

Two potential explanations may be given as to why radial cracks are not visible in tomography scanning of specimens [15]. The first one is that they trigger the main load drop, and may facilitate the removal of a larger part of material loaded in tension. This means that they are essentially removed during crater formation corresponding to the major load drop. Another explanation is that these cracks are closed when scanning post-mortem samples due to elastic unloading. This means that tomography should be performed while loading is applied, and the cracks are open or alternatively for other specimen geometries.

The latter case was investigated with a smaller cylindrical specimen (45 mm in diameter, 40 mm in height) that was indented in situ to ensure that splitting of the specimen occurred before the major load drop observed when a large granite sample is indented. In such situation, material removal would not hide radial cracking close to the indenter. Otherwise, the experiment was exactly the same as those performed on large granite blocks [15]. The post-mortem reconstructed volume for this experiment is shown in Figure 11. The fracture pattern may suggest that the failure of the entire specimen was due to the propagation of one of the radial cracks predicted by finite element simulations. This observation suggests that radial cracking occurred prior to complete failure (for smaller specimens) and major load drop (in case of larger samples).

It should be mentioned that the crater contains both the cracked region (due to tensile fracture) together with the crushed material due to compression and hydrostatic pressure. The damage results in this work only correspond to tensile cracking and the removed material due to crushing should also be included to properly capture the crater volume. Such an analysis can be performed by considering the high pressure region under the indenter as well as the shear zone to investigate the contribution of compressive inelasticity to the crater volume (Figure 12(a)). The average crater depth from the experiments corresponds to the region with high plastic deformation as shown in Figure 12(b). Under this depth, there is a part of the material that experiences plastic deformations without being fully damaged (removed). This comminuted zone corresponds to the region with multiple microcracks due to shear stresses, where irreversible strains are modeled as plasticity within the context of the current study. Statistical effects are as already stated above very limited.

It seems appropriate here to summarize the damage modeling for clarity. First, damage per se does not influence the load-displacement curve. This may be expected due to compressive loading during the indentation phase. However, a proper damage description is of utmost importance as it will give information about the cracking behavior in the contact region (and the drilling efficiency in practical situations), and also accurate values for local field variables. As regards to the last important feature related to model calibration, statistical damage effects are of less importance when the determination of the $m$-value reported in Table 1 is at issue as damage at first circumferential cracking was very limited. In short, the damage pattern is not much affected even though the first load drop is of statistical nature and is totally controlled by the Weibull parameters. The damage model is based on the fact that both tensile strength and effective volume in the DFH model are determined for a situation more closely related to the contact configuration. In this context, it should be mentioned that a similar approach was followed [30] but for another damage pattern. It should be remembered that, as discussed earlier, the complete set of experimental results reported in Figure 3(a) suggest an $m$-value equal to 6. Performing a simulation with $m = 6$ will again result in good agreement with the experimental force-penetration curve, but will give a very limited damage region as shown in Figure 13. Therefore, based on the damage pattern observed in Figure 7 and Figure 13, it is possible to state that the set of parameters listed in Table 1 yields a good description of granite in indentation situations.

Another matter of interest concerns the prediction of the final (large) load drop. Such catastrophic failure involves substantial fragmentation of the material, which is hard to predict numerically using the present continuum modeling approach.

As a final comment, it should be mentioned that the plastic region is confined to a small zone beneath the indenter. This effect was discussed in the context of cavity expansion modeling of contact, as first suggested by Johnson [16] and later by Alehossein et al. [17], which is relevant for rock materials.

It seems appropriate as a final effort to investigate the accuracy of the present numerical approach by comparing the achieved results with the theoretical predictions based on the cavity expansion model [17]. Figure 3 in Ref. [17] is used for comparison purposes where the scaled radius of the plastic zone (with contact radius) as a function of scaled indentation depth (with sphere radius) is shown for different values on the parameter κ, as defined in Eq. (11) of Ref. [17]. This parameter takes the value 0.0036 (with the shear modulus μ computed from Table 1, and the uniaxial compressive strength from Ref. [10]) and in order to ensure elastoplastic contact an indentation depth 0.2 mm is investigated. Accordingly, a value approximately equal to 3.9 was reported for the scaled plastic zone size. The present numerical approach, Figure 14, predicts a value of 4.0. These two values are very close and give confidence in the present numerical simulations as well as the theoretical predictions [17].

Specifically, at early stages of indentation loading, the region with positive tensile stress levels is completely separate from the plastic region. Therefore, the material experiencing positive tensile stresses can be considered as the zone where the material behavior is purely elastic. This means that the assumption of the Weibull theory, which states that failure occurs as soon as the initiation of a macroscopic fracture from a microcrack, is valid. On the other hand, using a purely elastic solution for the whole material leads to unrealistic (i.e., high) force levels and thus higher stresses prior to first load drop occurrence.

# 7  Concluding remarks

The fragmentation of Bohus granite under quasi-static indentation loading was numerically investigated. Previously modified elastoplastic constitutive model was considered. The "DP" model accounts for inelastic deformations except tensile damage. In this study, it was attempted to capture the fragmentation process by combining a damage description (i.e., DFH model) with the DP model. The resulting constitutive model is called DP-DFH. A simulation using the parameters gathered in Table 1 yielded predictions that were close to experimental findings both in terms of force-penetration response and damage pattern.

It was suggested that the indentation test results by themselves may be used to calibrate the Weibull parameters at small effective scales such as indentation applications. This is due to the fact that it is not practical to perform other conventional tests such as 3-point bending at such small scales leading to similar effective volumes. The initial fragmentation, which corresponds to the first small load drop in the force-penetration curve, is likely due to high radial tensile stresses at or close to the contact boundary. This stress level is considered as the tensile strength of the material in indentation tests. Further, an advanced elastoplastic model, including dilatational effects, coupled with an anisotropic damage model was needed to fully capture the behavior of the granite material at issue. The radius of the formed crater on indented specimens was in good agreement with the numerical simulation results. The indentation test thus is an important source of information for such characterization.

Another matter of interest concerns the prediction of the final (large) load drop. Such catastrophic failure induces substantial fragmentation of the material, which is hard to predict from a numerical standpoint. Qualitatively, it can be stated that the radial cracks visible in Figure 7 triggered the final load drop. Further fine-tuning of the model is needed for a quantitative prediction of the final load drop. This task is left for future studies.

**Acknowledgements**

The authors would like to thank Dr. Erik Olsson for reading the manuscript and giving valuable comments.

31. Fischer-Cripps AC, Gloyna EF, Hart WH. *Introduction to contact mechanics*. Springer; 2000.19

**Table and table legends**

Table 1. Material parameters used in the numerical simulations with the DP-DFH model.

| Material parameter | |
|---|---|
| $E$ (GPa) | 52 |
| $\nu$ | 0.25 |
| $\rho$ (kg/m³) | 2630 |
| $\beta$ (º) | 51.7 |
| $d$ (MPa) | 153.3 |
| $\Psi$ (º) | Figure 1 |
| $m$ | 24 |
| $\sigma_w$ (MPa) | 120 |
| $Z_{eff}$ (mm³) | 1 |

**Figure legends**

Figure 1. Dilation angle ($\Psi$) as a function of applied pressure ($p$).

Figure 2. Schematic illustration of an indentation test where $P$ is the contact force, $h$ the penetration depth, $a$ the contact radius and $R$ the radius of the indenter.

Figure 3. Weibull plot to obtain the Weibull modulus $m$ from indentation test results, **(a)** considering all data points ($m = 6$), and **(b)** discarding the three data points showing very low tensile strengths ($m = 24$).

Figure 4. Finite element mesh. **(a)** Geometry of the rock specimen and rigid indenter. **(b)** 3D view. By means of symmetries, only one quarter of the problem was modeled.

Figure 5. Schematic flowchart of the numerical implementation.

Figure 6. Force-penetration (*P-h*) response from numerical simulations using different material models as well as the experimentally obtained response.

Figure 7. **(a)** Damage variable $\omega_1$ corresponding to a penetration depth of 0.5 mm from finite element simulation of an indentation test using the parameters listed in Table 1. **(b)** Damaged elements.

Figure 8. **(a)** Indented surface of two of the rock samples loaded up to the large load drop. **(b)** Predicted damage pattern.

Figure 9. Stress fields from the finite element simulation using the plasticity model without any damage description. The penetration depth is 0.5 mm. **(a)** Maximum principal stress field (only tensile stresses). **(b)** Corresponding direction for the region enclosed in the dashed box. **(c)** Mid-principal stress field (only tensile stresses). **(d)** Corresponding direction for the region enclosed in the dashed box. **(e)** Minimum principal stress field (compressive stresses). **(f)** corresponding direction for the region enclosed in the dashed box.

Figure 10. **(a)** Predicted damage pattern from 10 simulations. Only the results for smallest and largest damage zones are shown. **(b)** Associated equivalent plastic strain fields. Only the results for the smallest and largest plastic zones are shown.

Figure 11. 3D scan of a splitting crack in an indented cylindrical specimen.

Figure 12. **(a)** Hydrostatic pressure field from the finite element simulation using a plasticity model without any damage description. **(b)** Corresponding equivalent plastic strain. The penetration depth is 0.5 mm.

Figure 13. Damage variable $\omega_1$ corresponding to a penetration depth of 0.5 mm from finite element simulations using a Weibull modulus $m = 6$.

Figure 14. Equivalent plastic strain field at a penetration depth of 0.2 mm determined from finite element simulations. The contact radius is 1.4 mm.

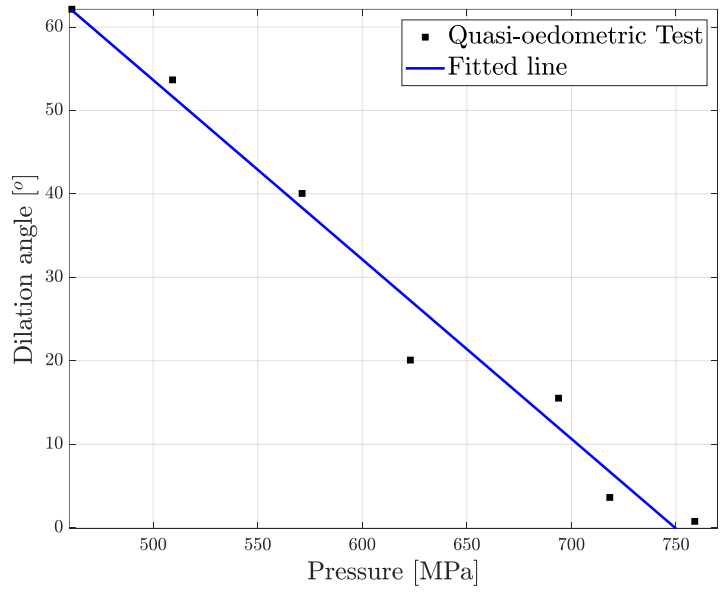

Figure 1

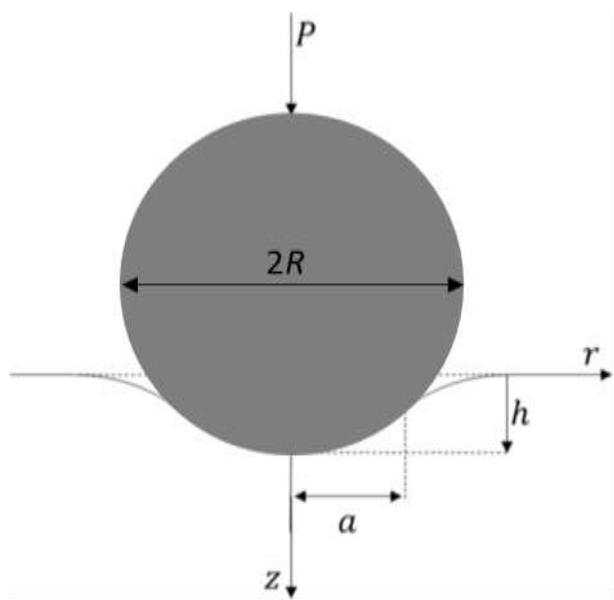

Figure 2

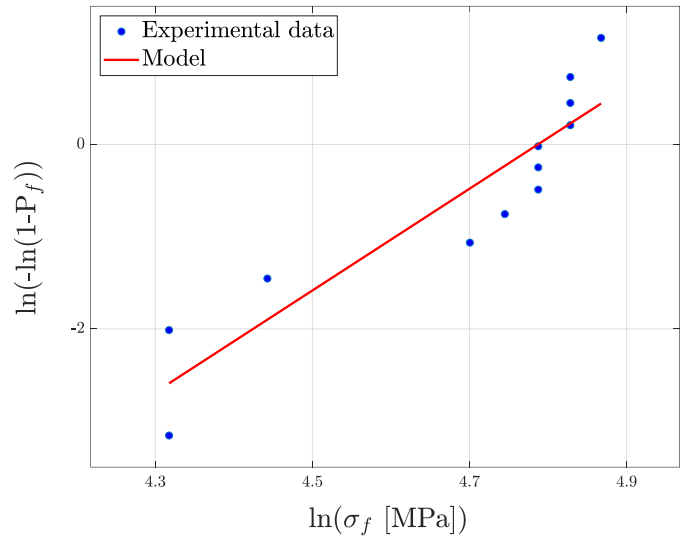

(a)

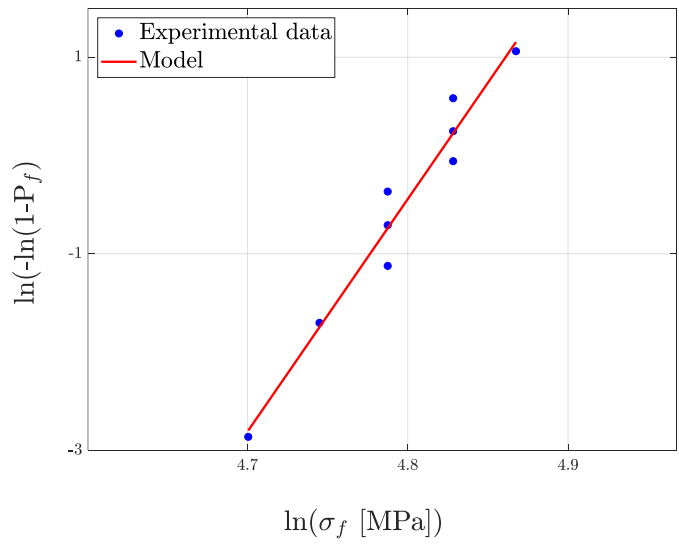

(b)

Figure 3

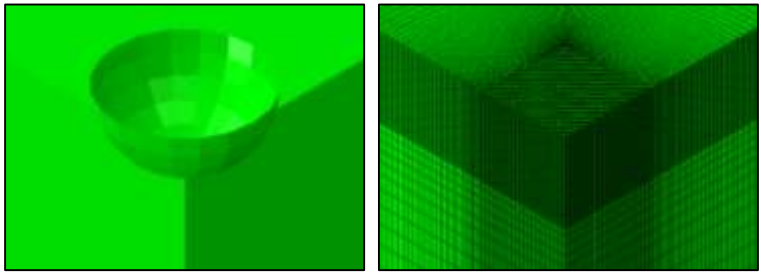

(a) (b)

Figure 4

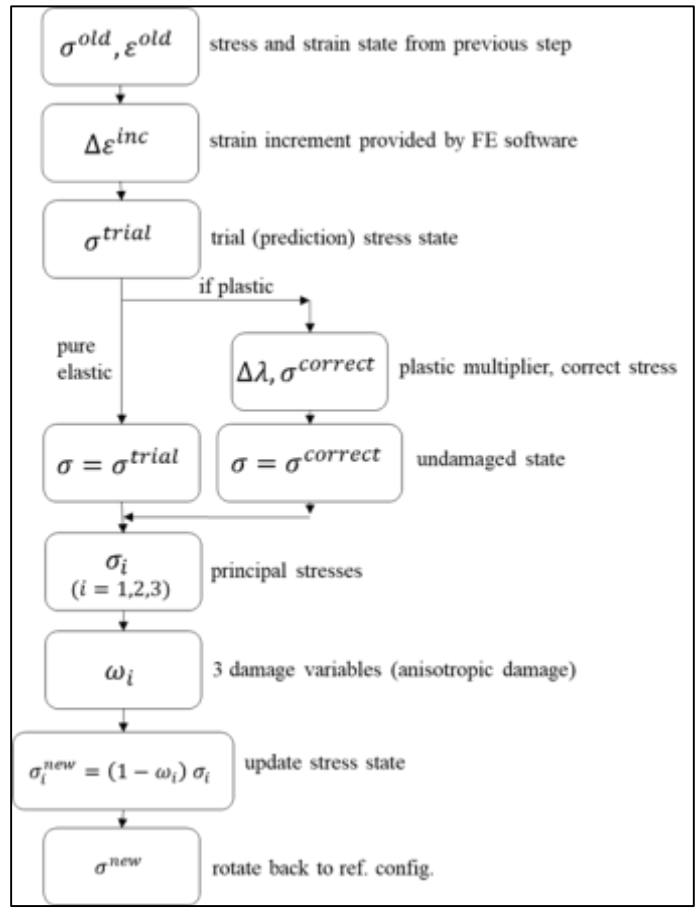

Figure 5

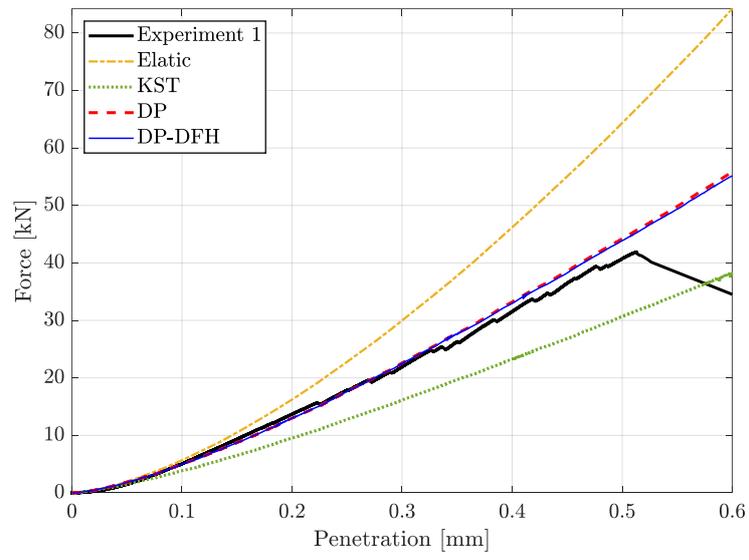

Figure 6

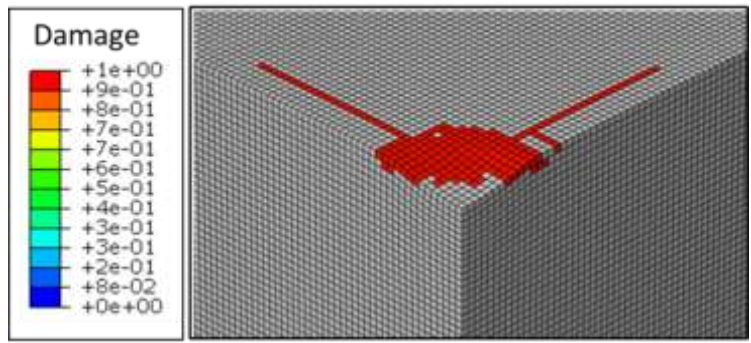

(a)

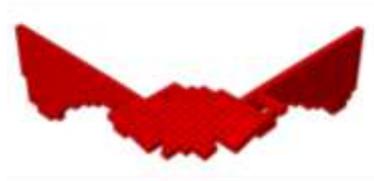

(b)

Figure 7

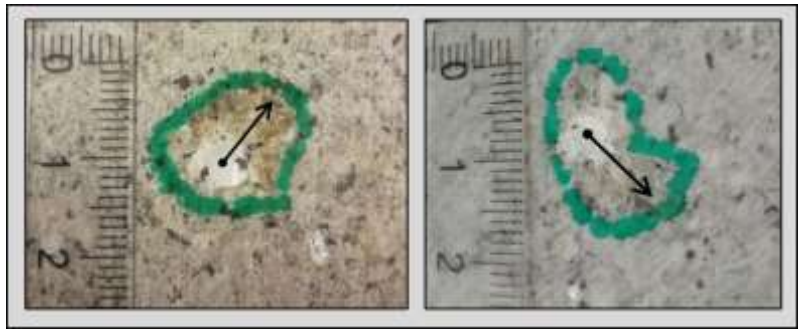

(a)

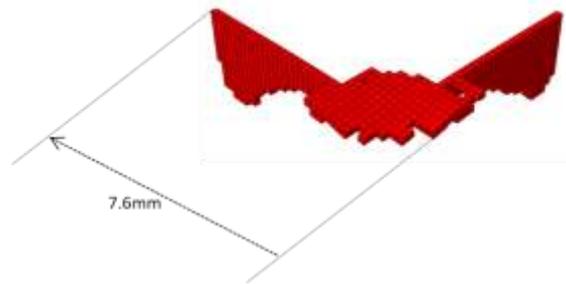

(b)

Figure 8

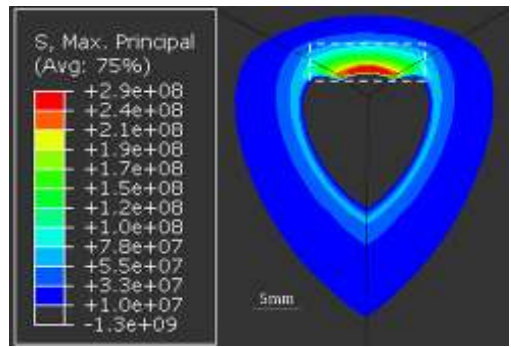

(a)

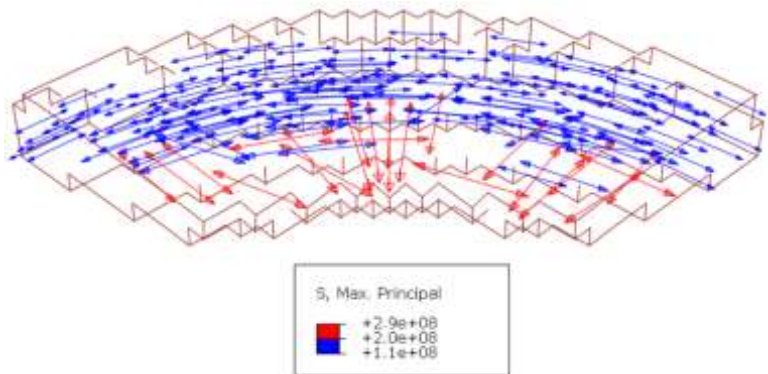

(b)

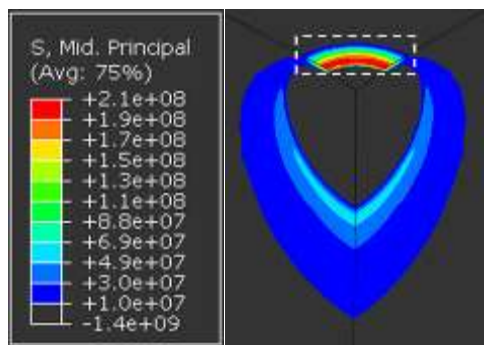

(c)

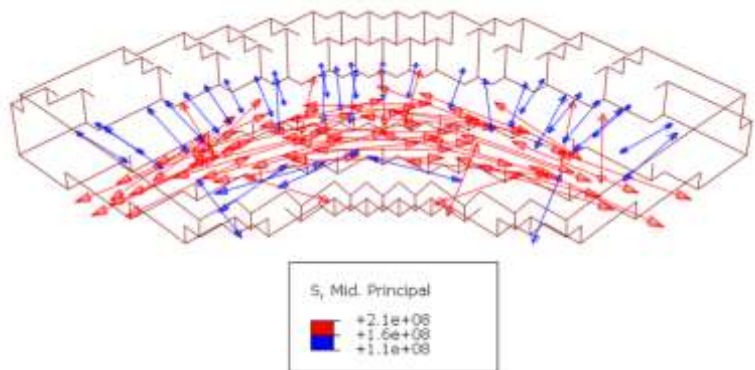

(d)

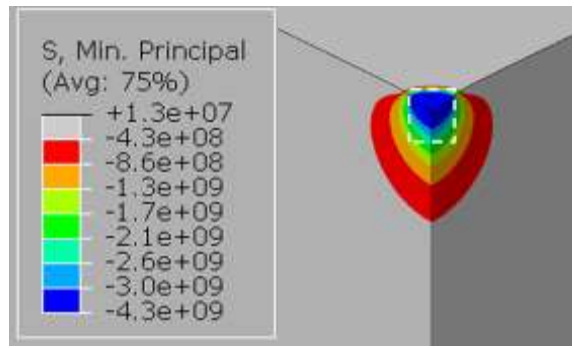

(e)

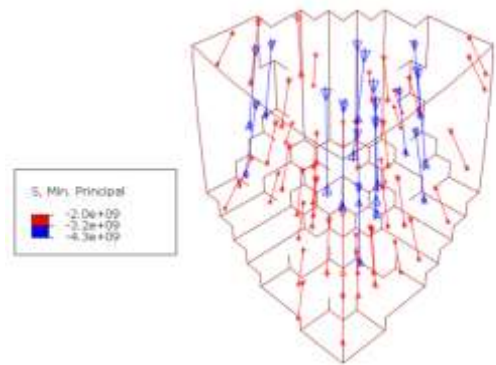

(f)

Figure 9

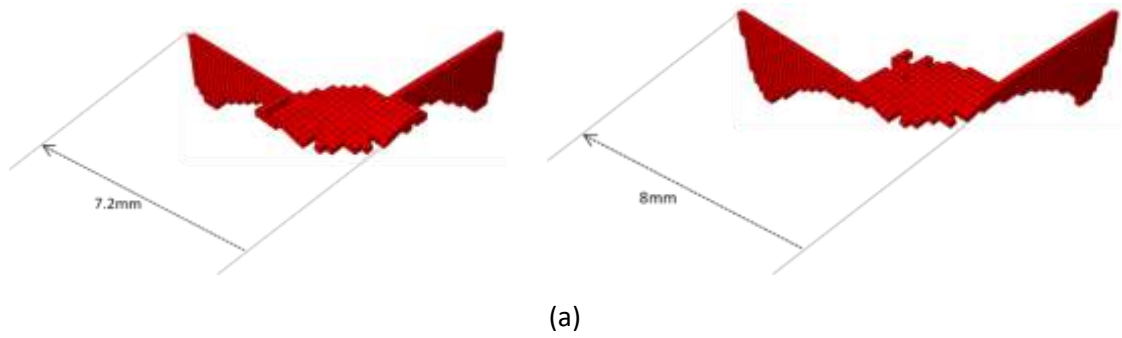

(a)

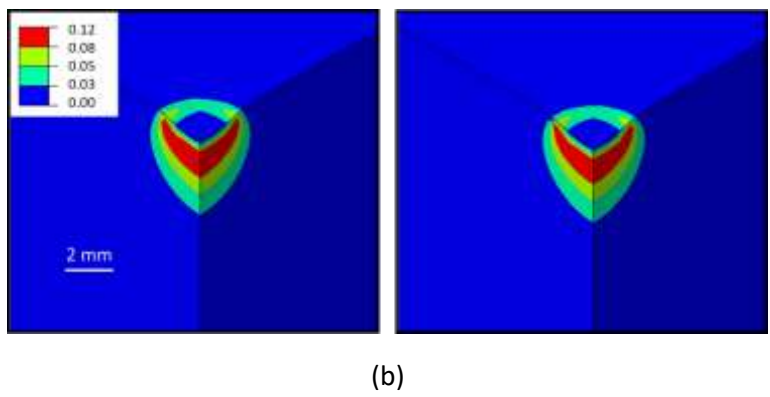

(b)

Figure 10

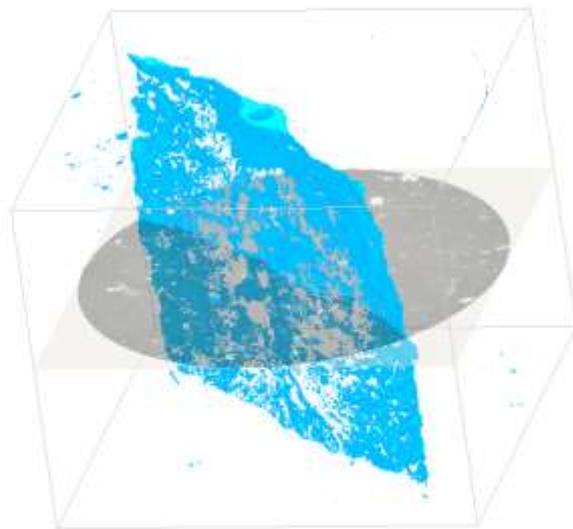

Figure 11

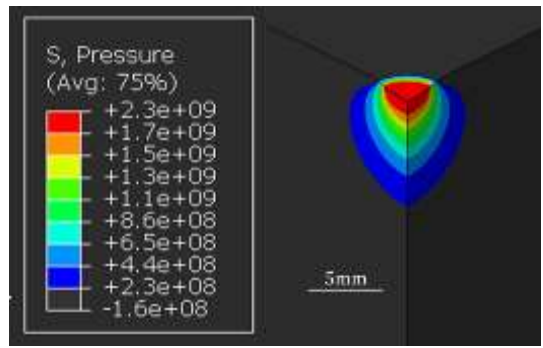

(a)

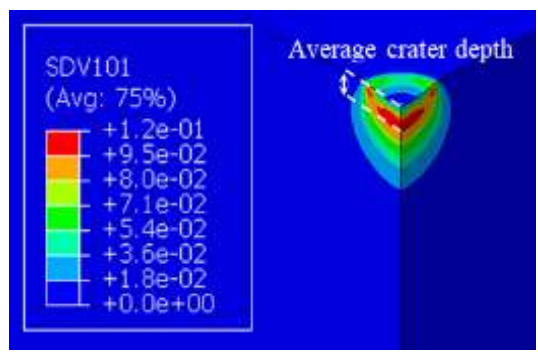

(b)

Figure 12

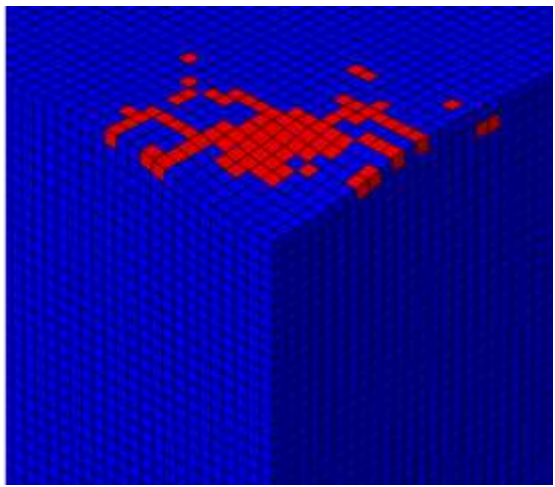

Figure 13

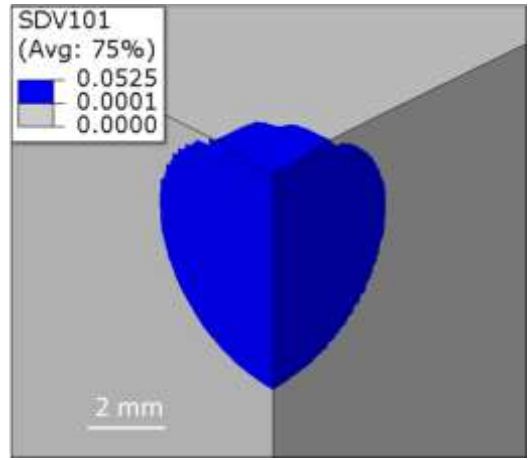

Figure 14